\documentclass[]{interact}

\usepackage{epstopdf}% To incorporate .eps illustrations using PDFLaTeX, etc.
\usepackage[caption=false]{subfig}% Support for small, `sub' figures and tables
\usepackage{float} 
\usepackage{url}
\usepackage{comment}
\usepackage{textcomp}
\usepackage{url}
\usepackage{color}

\usepackage[numbers,sort&compress]{natbib}% Citation support using natbib.sty
\bibpunct[, ]{[}{]}{,}{n}{,}{,}% Citation support using natbib.sty
\renewcommand\bibfont{\fontsize{10}{12}\selectfont}% Bibliography support using natbib.sty
\makeatletter% @ becomes a letter
\def\NAT@def@citea{\def\@citea{\NAT@separator}}% Suppress spaces between citations using natbib.sty
\makeatother% @ becomes a symbol again

\theoremstyle{plain}% Theorem-like structures provided by amsthm.sty

\theoremstyle{definition}

\theoremstyle{remark}

\begin{document}
%\articletype{Research Article}% Specify the article type or omit as appropriate

\title{Quantitative Prediction of Fracture Toughness ($K_{{\rm I}c}$) of Polymer
by Fractography Using Deep Neural Networks}

\author{
\name{Y. Mototake\textsuperscript{a}, K. Ito\textsuperscript{b}, and M. Demura\textsuperscript{b}}
\affil{\textsuperscript{a}The Institute of Statistical Mathematics, Tachikawa, Tokyo 190-8562, Japan. \textsuperscript{b}Research and Services Division of Materials Data and Integrated System, National Institute for Materials Science, 1-1 Namiki, Tsukuba, Ibaraki 305-0044, Japan.}
}

\maketitle
\begin{abstract}
Fracture surfaces provide various types of information about fracture.
The fracture toughness $K_{{\rm I}c}$, which represents the resistance to fracture, can be estimated using the three-dimensional (3D) information of a fracture surface, i.e., its roughness.
However, this is time-consuming and expensive to obtain the 3D information of a fracture surface; thus, it is desirable to estimate $K_{{\rm I}c}$ from a two-dimensional (2D) image, which can be easily obtained.
In recent years, methods of estimating a 3D structure from its 2D image using deep learning have been rapidly developed.
In this study, we propose a framework for fractography that directly estimates $K_{{\rm I}c}$ from a 2D fracture surface image using deep neural networks (DNNs).
Typically, image recognition using a DNN requires a tremendous amount of image data, which is difficult to acquire for fractography owing to the high experimental cost. 
To compensate for the limited data, in this study, we used the transfer learning (TL) method, and constructed high-performance prediction models even with a small dataset by transferring machine learning models trained using other large datasets. 
We found that the regression model obtained using our proposed framework can predict $K_{{\rm I}c}$ in the range of approximately 1--5 [MPa$\sqrt{m}$] with a standard deviation of the estimation error of approximately $\pm$0.37 [MPa$\sqrt{m}$]. 
The present results demonstrate that the DNN trained with TL opens a new route for quantitative fractography by which parameters of fracture process can be estimated from a fracture surface even with a small dataset. 
The proposed framework also enables the building of regression models in a few hours. 
Therefore, our framework enables us to screen a large number of image datasets available in the field of materials science and find candidates that are worth expensive machine learning analysis.
\end{abstract}

\section{Introduction}
Fracture surfaces provide various types of information on fracture.
For example, a dimple-like pattern indicates the occurrence of plastic deformation during fracture and a consequent requirement of a large amount of energy to fracture the material\cite{suresh1998fatigue}.
However, smooth fracture surfaces as seen in cleavage fractures are suggestive of brittle fracture\cite{suresh1998fatigue}. 
From a fracture surface, quantitative parameters of the fracture process, such as the fracture toughness $K_{{\rm I}c}$ or Charpy impact absorption, can be estimated.
In previous studies, such parameters have been estimated using the three-dimensional (3D) information of a fracture surface, i.e., its roughness.
For example, Mandelbrot et al. obtained the 3D landscape of a fracture surface by observing its structural changes while continuously slicing the sample. 
They found a correspondence between the Charpy impact absorption and the fractal dimension by the fractal analysis of this contour structure\cite{mandelbrot1984fractal}.
Another study revealed that $K_{{\rm I}c}$ can be estimated by extracting the roughness of a fracture surface measured by stereo imaging using an electron microscope\cite{mandelbrot1984fractal, barak2019correlating}.
From an engineering perspective, it is time-consuming and expensive to obtain the 3D information of a fracture surface; thus, it is desirable to estimate parameters of fracture process such as $K_{{\rm I}c}$ from two-dimensional (2D) images, which can be acquired at a relatively low cost.
The results of some previous studies have suggested that the contrast in 2D images contains 3D roughness information and that the identification of the fracture type, such as ductile or brittle fracture, is possible\cite{suresh1998fatigue, Underwood1986, bastidas2016fractographic}.
However, this contrast is also affected by the angle of the measurement target, the type of observation probe (such as light or electrons), and the interaction between the probe and the measured sample.
Therefore, in general, extracting a 3D structure from contrast is difficult in principle.
Thus, compared with previous methods that rely on 3D information, estimating parameters of fracture process such as $K_{{\rm I}c}$ from 2D images is difficult. \par
In this study, we establish a framework for fractography by which we can directly estimate $K_{{\rm I}c}$ from 2D fracture surface images using a deep neural network (DNN) by the transfer learning (TL) method.
In recent years, deep learning methods to estimate the depth position of an object from a single 2D image have been rapidly developed\cite{khan2020deep, alhashim2018high}. 
Thus, deep learning can extract the features of an uneven structure from the contrast of its 2D images under the effects of the above-mentioned complex factors.
These recent advances in deep learning research suggest that $K_{{\rm I}c}$ can be directly estimated from a 2D fracture surface image by extracting its features through deep learning.
Typically, image recognition using a DNN requires a tremendous amount of image data, but it is difficult to acquire such a large amount of image data for fractography owing to the high cost of experimentation involving specimen preparation, observation, and the measurement of fracture properties.
The dataset utilized in this study contains 770 2D fracture surface images with corresponding $K_{{\rm I}c}$; this is a large number for such types of data but quite small as a training dataset for a regular DNN, as subsequently described in detail.
To compensate for the limited data, we used the TL method, which allowed us to construct high-performance prediction models even with a small dataset by transferring machine learning models trained using other large datasets.
TL can be used to identify fracture surfaces using DNNs\cite{bastidas2020deep, tsopanidis2020toward, bastidas2019textural, konovalenko2018investigation, tsopanidisunsupervised}. 
Thus, it is expected that by using a DNN and TL, a framework for directly estimating $K_{{\rm I}c}$ can be established.
Concretely, in this study, we examined whether a DNN can extract the features required for predicting $K_{{\rm I}c}$ from images of fracture surfaces of polymers. 
The present fracture dataset of polymers, which was collected by the National Institute of Technology and Evaluation\cite{nite}, includes 770 data items, each being a set of one fracture surface image and its $K_{{\rm I}c}$. 
This dataset contains macroscale fracture surface data generated by various fracture processes, such as brittle and ductile fracture processes.
The results obtained in this study showed that even with such limited data, $K_{{\rm I}c}$ can be estimated from 2D fracture surface images using the established DNN and TL framework. \par
The remainder of this paper is organized as follows.
First, in Sec. II, we describe the fracture toughness test data of the polymeric materials used in this study.
In Sec. III, the DNN model used for TL and the regression model are explained.
In Sec. IV, the results of the analysis are presented, and in Sec. V, the results are discussed and summarized.

\section{Database of polymer fracture toughness test }
\begin{figure}[t]
  \begin{center}
   \includegraphics[width=80mm]{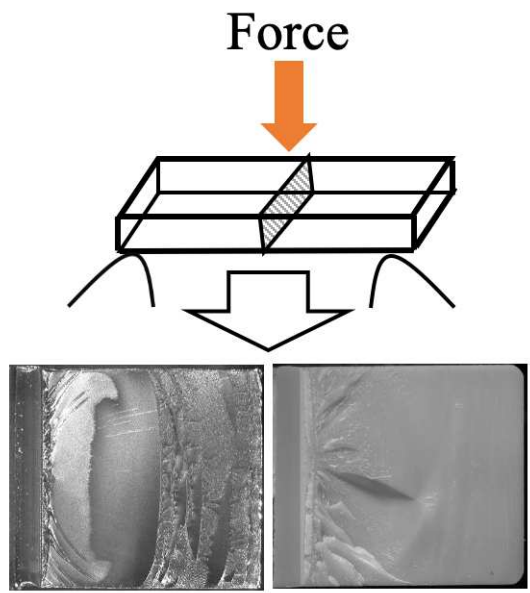}
  \caption{Upper figure: schematic of fracture toughness test. A notched sample is fractured by adding force, and the fracture surface is observed.
Lower figures: examples of fracture surface image obtained in the test\cite{nite}.}
  \label{fig1}
  \end{center}
\end{figure}

In this study, the ``Polymer Fracture Database''\cite{nite} is used, which is provided by the National Institute of Technology and Evaluation (NITE). 
The development of this database as a collaborative effort of NITE, Yamagata University, Osaka Municipal Technical Research Institute, and Meiji University as a project commissioned by the New Energy and Industrial Technology Development Organization (NEDO) began in 1996. 
Since 2020, the database has been available on the Materials Data Repositry (MDR) managed by the National Institute for Materials Science (NIMS). \par
This database was established to develop a widely used technological infrastructure to enhance the safety of materials, such as the development of technologies for investigating the causes of accidents arising from the fracture or failure of plastic products. 
The database contains test data for various polymeric materials used in society, which were obtained from the fracture toughness, tensile, fatigue, creep, essential work of fracture (EWF), and scratch tests. 
In this paper, data from the fracture toughness test were used, because such data (upper figure of Fig.~\ref{fig1}) are the most comprehensive. 
For the fracture toughness test, data were obtained for various types of samples with different thicknesses, compositions, grades, molding methods, mold temperatures, injection speeds, cylinder temperatures, and with/without welds, at different test temperatures and speeds. 
The test temperatures were centered at room temperature and generally ranged from $-40$ \textdegree{}C to $+60$ \textdegree{}C. 
The database contains macro-photographs taken of about a 10 mm square of the fracture surface for each set of the above conditions (lower figures of Fig.~\ref{fig1}). 
One fracture surface image was included for each of the 1248 test conditions. 
Among the parameters in the database that are expected to be estimable, as mentioned in Introduction, $K_{{\rm I}c}$ for plane strain was selected in this study because it is the main parameter that one would aim to obtain in a fracture toughness test, and it has always been recorded in many tests in the database. \par
The database includes comma separated values (CSV) files listing information for all tests, PDF files containing fracture surface photos for each set of test conditions, and tab separated values (TSV) files containing load-displacement curve data. 
Each test was assigned an identification code. 
Since the fracture surface images were recorded in the lossy compressed JPEG format in the PDF file of each test and bitmap data without degradation could not be obtained, the image data were extracted, as it is in the JPEG format, using the pdfimages\cite{pdfimages} command. 
Note that the extracted image file had its left and right sides reversed from when the PDF was displayed, so the left and right sides were reversed using the convert command of Imagemagick\cite{imagemagic}. 
From among all test data, 770 data records were obtained by extracting test data in which $K_{{\rm I}c}$ was recorded and which included fracture surface data.
These 770 data records were used as the dataset for the analysis in this study. 

\section{Method}
\begin{figure}[htbp]
  \begin{center}
   \includegraphics[width=150mm]{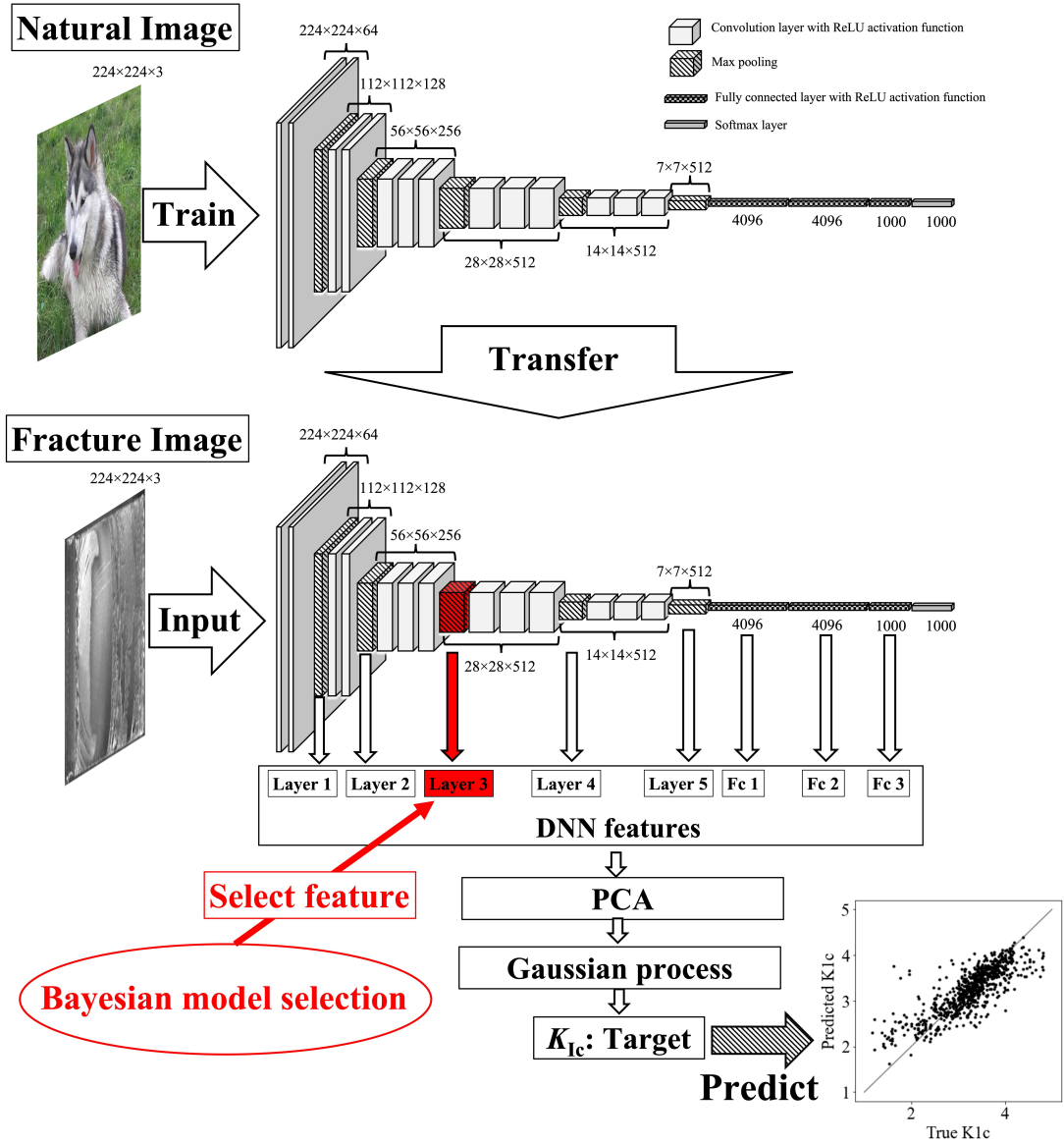}
  \caption{Conceptual diagram of the proposed transfer learning framework for estimating $K_{{\rm I}c}$.}
  \label{fig2}
  \end{center}
\end{figure}

In this section, we describe the method adopted for building a predictive model of $K_{{\rm I}c}$ using feature-extraction-type transfer learning\cite{li2017learning}, which is a combination of a DNN trained on natural images and Gaussian process regression\cite{williams2006gaussian}. 
Specifically, we developed a model that uses a DNN as a feature extractor and regresses its features and $K_{{\rm I}c}$ by Gaussian process regression. \par
The DNN model used in this study was the vgg16 model\cite{simonyan2014very}, which identifies natural images such as animals and vehicles. 
vgg16 is composed of convolution, pooling, and fully connected layers.
A convolution layer convolves the input space using a convolution filter.
Convolution filters of DNNs trained on natural images frequently behave similarly to Gabor filters. 
A Gabor filter can extract image features such as the edges of graphic structures in images. 
In vgg16, a structure comprising a stack of convolution layers 
alternated with pooling layers is formed, and the pooling layers compress the dimensions by replacing certain regions of the convolution layers with representative values such as the maximum and average of the region. 
Pooling can be viewed as a transformation that renders the input space coarse grained. 
Thus, after training on natural images, vgg16 is expected to extract useful features on various scales for natural images. 
In the prediction of $K_{{\rm I}c}$ using fracture surface images, the evaluation of the edges of the graphic structures in the images is also considered important. 
Depth estimation research using images has shown that using a DNN pretrained with the natural image dataset called ImageNet\cite{deng2009imagenet} as a feature extractor improves the accuracy of depth estimation after fine-tuning, compared with the method without pretraining\cite{alhashim2018high}. 
This finding suggests that DNNs trained on Imagenet can extract the image features necessary for depth estimation. 
As noted in Introduction, information on depth is useful in predicting $K_{{\rm I}c}$. 
Thus, the image features extracted by vgg16 trained on natural images at each convolution layer are considered to yield useful features for predicting $K_{{\rm I}c}$. 
Therefore, we employed feature-extraction-type transfer learning\cite{li2017learning}, that used vgg16 trained on natural images as an extractor of image features to construct a prediction model of $K_{{\rm I}c}$. 
With such transfer learning, a framework can be constructed for learning a large amount of data at a lower cost owing to the exclusion of the fine-tuning of the DNN. 
However, in the absence of fine-tuning, if certain image features required for the prediction of $K_{{\rm I}c}$ are not included in the set of image features required for the classification of Imagenet, then they remain missing. 
In this study, we examined whether the prediction of $K_{{\rm I}c}$ is still possible under such conditions.
\par
The predictive model of $K_{{\rm I}c}$ was constructed as follows. 
First, we prepared vgg16 (top figure of Fig.~\ref{fig2}) trained on the ImageNet dataset.
In this study, publicly available learned weights\cite{vgg16weight} were used as the trained vgg16. 
In addition, a library\cite{david} using TensorFlow was adopted as a prototype for the vgg16 implementation. 
When an image is input to the pretrained vgg16, the features are obtained as the outputs of the five pooling and three fully connected layers referred to as Layer 1--Layer 5 and Fc 1--Fc 3, respectively. 
We denote one of these DNN features as $m\in\{{\rm Layer\:1,Layer\:2,Layer\:3,Layer\:4,Layer\:5,Fc\:1,Fc\:2,Fc\:3}\}$. 
Let $\mathbf{h}^i_m$ be a $d_m$-dimensional DNN feature for a certain fracture image $i$. 
Because the number of data $N=770$ was much smaller than $d_m$, $\mathbf{h}^i_m$ was compressed to $N$ dimensions by principal component analysis (PCA) to reduce the computational complexity. 
Let $\mathbf{x}^i_m$ be a compressed DNN feature. 
Note that in a PCA that compresses data into the dimension of the sample size, information on the covariance of the dataset is not lost. 
Therefore, a regression model should be learned to predict $K_{{\rm I}c}$ perfectly consistent with that in the uncompressed case in Gaussian process regression. 
Each fracture surface image, which was approximately $400 \times 400$ pixels in size, was prescaled using zero-order spline interpolation\cite{scikit-image} to ensure that it had the same number of pixels as the natural image ($224\times224$).
A predictive model of $K_{{\rm I}c}$ was constructed by Gaussian process regression using the dimension-compressed DNN feature $\mathbf{x}^i_m$ as the explanatory variable (bottom figure of Fig.~\ref{fig2}). 
Note that in the entire regression model consisting of vgg16 and Gaussian process regression, only the regression parameters of the latter were trained on the fracture surface image data, and the DNN model used the parameters from the pretraining as is. 
Gaussian process regression allows the flexible construction of regression models by appropriately choosing a kernel $k(\cdot,\cdot)$. 
In this study, the rational quadratic kernel
\begin{eqnarray}
k(\mathbf{x}^i_m, \mathbf{x}^j_m) := \theta_0^2\left(1+\frac{\left|\mathbf{x}^i_m - \mathbf{x}^j_m\right|_2^2}{2\theta_1 \theta_2^2}\right)^{-\theta_1}
\end{eqnarray}
and the Gaussian kernel
\begin{eqnarray}
\phi(\mathbf{x}_n,\mathbf{x}_m) = \theta_0^2\exp\left[-\frac{1}{2\theta_1^2}\left\lVert\mathbf{x}_n-\mathbf{x}_m\right\rVert^2\right] 
\end{eqnarray}
were the candidates for $k(\cdot,\cdot)$.
Here, $\theta_k$, which determines the shape of the kernel, is the hyperparameters of the Gaussian process regression. \par
To construct such a regression model, the DNN feature $m$, the kernel function type $k$, and the kernel function hyperparameters $\theta_k$ must be properly selected. 
In this study, these parameters are selected on the basis of a Bayesian inference framework\cite{bishop2006pattern}. 
For the selection of $m$ and $k$, the Bayesian model selection framework is employed. 
On the basis of an indicator called Bayesian free energy\cite{watanabe2018mathematical}, the Bayesian model selection framework selects the model that considers discrete-valued hyperparameters such as $m$ or $k$ (see Appendix A). 
In general, when using machine learning to build predictive models, it is necessary to prevent over-fitting, where the model over-fits the training data and loses predictive performance. 
Using Bayesian free energy as an indicator for model selection, it is possible to select a model in which over-fitting does not occur. 
$\theta_k$, that is, the continuous-valued hyperparameter, can also be estimated on the basis of Bayesian free energy (see Appendix A). 
Thus, the model and hyperparameters are selected so as to have a lower Bayesian free energy. 
$\theta_k$ was optimized using the L-BFGS-B algorithm~\cite{liu1989limited}, a type of quasi-Newtonian method. 
Newton's method sometimes gives a local solution depending on the initial values of the parameters to be optimized. 
Therefore, in this study, we searched for initial values that minimize the Bayesian free energy in the range shown in Table~\ref{tbl1}. 
The brackets $(\:,\:)$ to the right of each parameter in Table~\ref{tbl1} indicate how to set the grid. 
The first element in the brackets is a scale of the space to be gridded and the second element is the number of grids.\par
For comparison with the case where the DNN feature is not used, a regression of $K_{{\rm I}c}$ was performed directly from the fracture surface images by Gaussian process regression with the same procedure as in the case of DNN features. 
Hereafter, the obtained regression model is referred to as the ``direct regression'' model.
\begin{table}[t]
\centering
 \caption{Search range for hyperparameters of kernel}
        \begin{tabular}{|c|c|c||c|c|}\hline
        \multicolumn{3}{|c||}{Rational quadratic kernel} & \multicolumn{2}{|c|}{Gaussian kernel}\\ \hline
            $\theta_0$(linear, 5 grids)& $\theta_1$(log, 5 grids)& $\theta_2$(log, 10 grids) & $\theta_0$(linear, 20 grids) & $\theta_1$(log, 20 grids) \\ \hline
            0.01--0.2 & $10^{-2.0}$--$10^{2.0}$ & $10^{1.0}$--$10^{1.2}$ & 0.01--0.2 & $10^{1.0}$--$10^{3.0}$  \\ \hline
        \end{tabular}
        \label{tbl1}
\end{table}

\section{Results and Discussion}
\begin{figure}[b]
  \begin{center}
   \includegraphics[width=140mm]{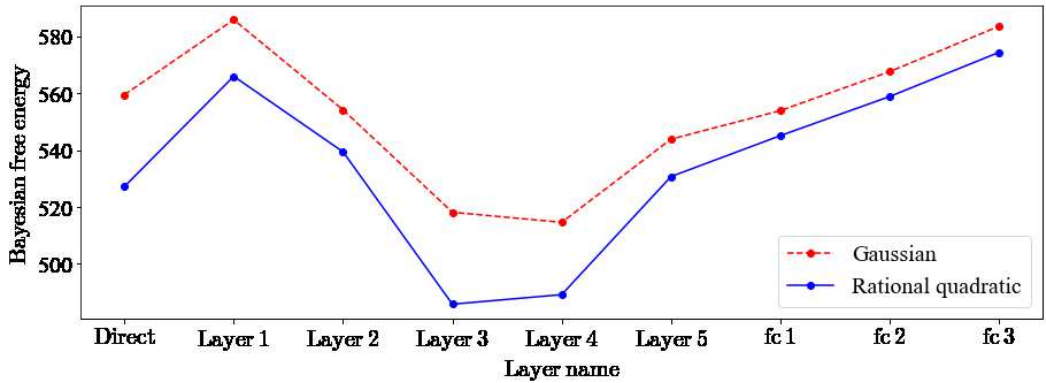}
  \caption{Results of Bayesian model selection for DNN features and kernel types.}
  \label{fig3}
  \end{center}
\end{figure}
As a result of Bayesian model selection, Layer 3 with a rational quadratic kernel is the optimal model for predicting $K_{{\rm I}c}$, where the Bayesian free energy is minimum (Fig. \ref{fig3}). 
For all DNN features, the Bayesian free energy with the rational quadratic kernel is always lower than that with the Gaussian kernel. 
This indicates that the rational quadratic kernel is always a plausible kernel type for any DNN feature. 
In terms of optimal DNN features, Layers 3 and 4 were confirmed to be plausible features specifically with a lower Bayesian free energy than the others, regardless of the kernel type. 
The most plausible regression model with the minimum Bayesian free energy was obtained when the rational quadratic kernel was selected as the kernel function and Layer 3 for the DNN features. 
The difference in the Bayesian free energy of the regression model using Layer 3 and Layer 4 features in the rational quadratic kernel was only about 3.36. 
However, since Bayesian free energy is defined as the logarithm of the likelihood function, which expresses the likelihood of the model (Appendix A), the features of Layer 3 are more than $\exp(3.36)=28$ times more likely than those of Layer 4 when compared in the original linear space. 
Therefore, a model that performs regression using a rational quadratic kernel with Layer 3 is selected here. 
This selected regression model has a lower Bayesian free energy than direct regression. 
It is noteworthy that under the same kernel function, there are DNN features that yield regression models with lower performance in terms of Bayesian free energy than direct regression. 
For example, in the rational quadratic kernel, only Layers 3 and 4 had a lower Bayesian free energy than direct regression, and other DNN features such as convolution and fc layers produced regression models with a higher Bayesian free energy than direct regression. 
Under a feature extraction type of transfer learning, higher-layer features than the last layer of the convolution layer\cite{li2017learning, deniz2018transfer, napel2018quantitative, kaur2020deep, gopalakrishnan2017deep}, Layer 5 in our model, are often used, but in the case of this study, they were not considered. 
In other words, depending on the material to be analyzed and the mechanical properties to be estimated, transfer learning would not be effective unless the appropriate Layer is selected. 
The selected Layer 3 corresponds to a resolution of about 10$\%$ of the input image. 
This means that the spatial scale of the best DNN feature is on a relatively coarse-grained scale. 
It is suggested that a complex feature, different from the simple Gabor filter, is useful for $K_{{\rm I}c}$ estimation.\par 
\begin{figure}[t]
  \begin{center}
   \includegraphics[width=140mm]{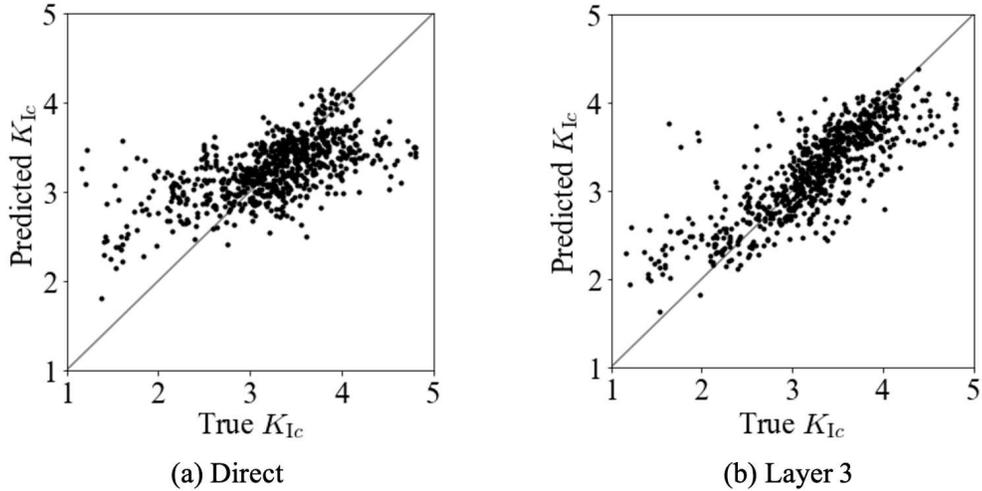}
  \caption{Integrated prediction results for test data in all test folds, which are data not used for training. 
  (a) Result of direct regression from fracture surface images using Gaussian process regression with rational quadratic kernel. 
  (b) Result of regression from Layer~3 feature using Gaussian process regression with rational quadratic kernel. 
  }
  \label{fig4}
  \end{center}
\end{figure}
The regression of $K_{{\rm I}c}$ was performed using the DNN features, the kernel type, and the kernel hyperparameters $\theta_k$ selected in terms of Bayesian free energy. 
Figure~\ref{fig4} shows the regression results. 
To use all the data effectively, we estimated the prediction performance on the basis of the framework of the 20-fold cross-validation method. 
The black dots are the integrated prediction results for test data in all test folds where data are not used for training. 
The horizontal axis is the measured value and the vertical axis is the predicted value of the constructed model. 
The left figure shows the result of predicting $K_{{\rm I}c}$ directly from the fracture image (direct regression model). 
The direct regression returns a constant of around 3 for most of the cases, and there is virtually no correlation between the predicted and measured values. 
This indicates that the direct regression method cannot extract enough information to predict $K_{{\rm I}c}$. 
Note that the regression model using DNN feature extraction with Layer 3 has high accuracy, as shown by Fig.~\ref{fig4}(b). 
The $R^2$ value of the prediction model using Layer 3 features was 0.65 (correlation coefficient is 0.81). 
We also found that our model can predict $K_{{\rm I}c}$ with a standard deviation of the estimation error of about $\pm$0.37 [MPa$\sqrt{m}$] in the range of around 1–5 [MPa$\sqrt{m}$] (Fig.~\ref{fig4}). 
The high accuracy demonstrates that it is possible to predict $K_{{\rm I}c}$ using only the 2D fracture image without 3D height information. 
The high accuracy also suggests that the morphology of the fracture surface contains some information about resistance to fracture.\par 
Thus, the comparison with the poor result of the direct prediction method proves that the transfer learning using a DNN trained on natural images as a feature extractor is crucial in building a predictor for $K_{{\rm I}c}$ that has high estimation accuracy. 
It was also revealed that the importance of the proper use of DNN information, i.e., the selective use of Layer 3 in this case, should be emphasized.

\section{Summary}
The present result demonstrates that the DNN trained with TL opens the way to quantitative fractography in which one can estimate the parameters of fracture process from the fracture surface image even with a small dataset. 
In this study, the DNN pretrained on the Imagenet dataset was used directly as a feature extractor. 
Fine-tuning the DNN with fracture surface image data is expected to improve the performance of the model obtained in this study. 
It is also assumed that a higher estimation accuracy of $K_{{\rm I}c}$ could be achieved by using not only the fracture surface image but also the additional explanatory variables describing the target material, such as the production conditions or the operating environment. \par
The proposed framework, which enables the construction of regression models in a few hours, can be used for the low-cost screening of profitable candidates on which to apply machine learning analysis to estimate values representing mechanical property from a large number of image datasets available in the field of materials science. 
Therefore, if the utilization of the proposed framework is expanded, it is expected that machine learning methods in materials science will be used more broadly, quickly, and efficiently. 

\section*{Acknowledgements}
This work was supported by the Council for Science, Technology and Innovation (CSTI), Cross-Ministerial Strategic Innovation Promotion Program (SIP), “Structural Materials for Innovation” and “Materials Integration for Revolutionary Design System of Structural Materials” (funding agency: JST), JST PRESTO Grant Number JPMJPR212A, and JSPS KAKENHI Grant-in-Aid for Scientific Research on Innovative Areas ``Discrete Geometry for Exploring Next-Generation Materials'' 20H04648 and 22K13979.

\section*{Appendix A Gaussian process regression and Bayesian free energy}
\label{appendixa}
Let $\mathbf{x}^i_m$ be the N-dimensional compressed output of the $m$-th layer using PCA when a certain fracture surface image $i$ is input to the pretrained DNN. 
Also, let the $K_{{\rm I}c}$ value of fracture surface image $i$ be $y^i$. 
Suppose there are $N$ samples of $\mathbf{x}^i_m$ and $y^i$. 
Let them form the matrix $\mathbf{X} := (\mathbf{x}^1_m, \mathbf{x}^2_m \cdots \mathbf{x}^N_m)$ and the vector $\mathbf{Y} := (y_1,y_2 \cdots y_N)^T$. \par
Unlike regression based on the maximum likelihood method, Gaussian process regression does not involve the estimation of parameters of the regression function, such as regression coefficients in linear regression. 
In Gaussian process regression, the regression function $\mathbf{F}:=\mathbf{F}(\mathbf{X})=(f(\mathbf{x}^1_m), f(\mathbf{x}^2_m) \cdots f(\mathbf{x }^N_m))$ itself is the target of estimation, and a likelihood function in which $\mathbf{F}$ is the probability variable is introduced as
\begin{eqnarray}
{\rm P}(\mathbf{Y}|\mathbf{F})=\frac{1}{Z}\exp\left[-\frac{1}{2}\left(\mathbf{Y} - \mathbf{F}\right)^T \beta \mathbf{I}\left(\mathbf{Y} - \mathbf{F}\right)\right],
\end{eqnarray}
where $Z$ is the normalization constant and $\mathbf{I}$ is the identity matrix. 
From Bayes' theorem $P(B|A)=\frac{P(A|B)P(B)}{P(A)}$, the regression function $\mathbf{F}$ is estimated as the posterior probability distribution ${\rm P}(\mathbf{F}|\mathbf{Y})$ as
\begin{eqnarray}
{\rm P}(\mathbf{F}|\mathbf{Y}) &=& \frac{{\rm P}(\mathbf{Y}|\mathbf{F}){\rm P}(\mathbf{F})}{{\rm P}(\mathbf{Y})}
\\ &=&\frac{\frac{1}{Z}\exp\left[-\frac{\beta}{2}\left(\mathbf{Y} - \mathbf{F}\right)^T \mathbf{I}\left(\mathbf{Y} - \mathbf{F}\right)\right] \frac{1}{Z_2}\exp\left(-\frac{1}{2}\mathbf{F}^T \mathbf{K}^{-1} \mathbf{F}\right)}{{\rm P}(\mathbf{Y})},\\
\mathbf{K} &:=& \mathbf{K}(\mathbf{X},\mathbf{X}) = \left(\begin{matrix}
k(\mathbf{x}^1_m, \mathbf{x}^1_m) & \cdots & k(\mathbf{x}^1_m, \mathbf{x}^i_m) & \cdots & k(\mathbf{x}^1_m, \mathbf{x}^{N}_m)\\
\vdots & \ddots &  &  & \vdots\\
k(\mathbf{x}^i_m, \mathbf{x}^1_m) &  & k(\mathbf{x}^i_m, \mathbf{x}^i_m) &  & k(\mathbf{x}^i_m, \mathbf{x}^{N}_m)\\
\vdots &  &  & \ddots & \vdots\\
k(\mathbf{x}^{N}_m, \mathbf{x}^1_m) & \cdots & k(\mathbf{x}^{N}_m, \mathbf{x}^{i}_m) & \cdots & k(\mathbf{x}^{N}_m, \mathbf{x}^{N}_m) \label{eq_post1}\\
\end{matrix}\right),
\end{eqnarray}
where the prior distribution ${\rm P}(\mathbf{F})$ is assumed to be an $N$-dimensional normal distribution with mean 0 and variance--covariance matrix $\mathbf{K}$, and $k(\mathbf{x}^i_m,\mathbf{x}^j_m)$ is a kernel function that indicates the similarity between $\mathbf{x}^i_m$ and $\mathbf{x}^j_m$. 
In this study, the rational quadratic kernel
\begin{eqnarray}
k(\mathbf{x}^i_m, \mathbf{x}^j_m) := \theta_0^2\left(1+\frac{\left|\mathbf{x}^i_m - \mathbf{x}^j_m\right|_2^2}{2\theta_1 \theta_2^2}\right)^{-\theta_1}
\end{eqnarray}
 and the Gaussian kernel
\begin{eqnarray}
k(\mathbf{x}_m,\mathbf{x}_m) = \exp\left[-\frac{1}{2\theta_1^2}\left\lVert\mathbf{x}_m-\mathbf{x}_m\right\rVert^2\right] + \theta_0
\end{eqnarray}
 are employed as candidates for the kernel function. 
The hyperparameters of each type of kernel are defined as $\theta_{\rm Rational} = \{\theta_0, \theta_1, \theta_2\}$ and $\theta_{\rm Gauss} = \{\theta_0, \theta_1\}$. 
By reorganizing Eq.~(\ref{eq_post1}), it is revealed that the posterior distribution is a normal distribution with respect to $\mathbf{F}$. 
\begin{eqnarray}
{\rm P}(\mathbf{F}|\mathbf{Y}) &=& \frac{1}{Z'}\exp\left\{-\frac{1}{2}\left[\mathbf{F} - \left(\mathbf{K} + \beta^{-1}\mathbf{I}\right)^{-1}\mathbf{K}\mathbf{Y}\right]^T \left[\mathbf{K}^{-1} + \beta I\right] \left[\mathbf{F} - \left(\mathbf{K} + \beta^{-1}\mathbf{I}\right)^{-1}\mathbf{K}\mathbf{Y}\right]\right\},\:\:\:\:\:\:
\label{eq_post2}
\end{eqnarray}
where $Z'$ is the normalization constant. 
This type of regression in which the prior and posterior distributions are normal distributions is referred to as Gaussian process regression. 
Using the obtained regression model ${\rm P}(\mathbf{F}|\mathbf{Y})$ [Eq.~(\ref{eq_post2})], the predicted value $y^{\rm new}$ at the unknown point $\mathbf{x}_m^{\rm new}$ is given as the probability ${\rm P}(y^{\rm new}|\mathbf{Y},m,k,\theta_k)$:
\begin{eqnarray}
{\rm P}(y^{\rm new}|\mathbf{Y},m,k,\theta_k) &=& \frac{\int_{-\infty}^{\infty}d\mathbf{F} {\rm P}(y_{\rm new}|\mathbf{F},\theta_k,m,k) {\rm P}(\mathbf{Y}|\mathbf{F},\theta_k,m,k) {\rm P}(\mathbf{F}|\theta_k,m,k) {\rm P}(\theta_k,m,k)}{{\rm P}(\mathbf{Y}|\theta_k,m,k)}\:\:\:\:\:\:\:\:\\
&\propto& \exp\left[-\frac{1}{2}\left(
y_{\rm new} 
- \Sigma_{22}^{-1} \Sigma_{21}^T \mathbf{Y}
\right)^T\Sigma_{22}(y_{\rm new} - \Sigma_{22}^{-1}\Sigma_{21}^T\mathbf{Y} )\right],\\
\Sigma_{22}&=&\left(\Lambda_{22}-\Lambda_{21}^T\Lambda_{11}^{-1}\Lambda_{21}\right)^{-1},\:\:\Sigma_{21} =  \left(\Lambda_{22}-\Lambda_{21}^T\Lambda_{11}^{-1}\Lambda_{21}\right)^{-1}\Lambda_{21}\Lambda_{11}^{-1},\\
\mathbf{\Lambda}&=&
\left(\begin{matrix}
\Lambda_{11} & \Lambda_{12}\\
\Lambda_{21} & \Lambda_{22}
\end{matrix}\right)
=
\left(\begin{matrix}
\boldsymbol{K}(\mathbf{X},\mathbf{X}) + \beta^{-1}\mathbf{I}& \boldsymbol{K}(\mathbf{X},\mathbf{x}^{\rm new}_m)\\
\boldsymbol{K}(\mathbf{x}^{\rm new}_m,\mathbf{X}) & k(\mathbf{x}^{\rm new}_m,\mathbf{x}^{\rm new}_m) + \beta^{-1}
\end{matrix}\right),\\
\boldsymbol{K}(\mathbf{x}^{\rm new}_m,\mathbf{X}) &=& \boldsymbol{K}(\mathbf{X},\mathbf{x}^{\rm new}_m)^T = \left(k(\mathbf{x}^{\rm new}_m, \mathbf{x}^1_m), \cdots , k(\mathbf{x}^{\rm new}_m, \mathbf{x}^{N}_m)\right),
\end{eqnarray}
where the prior distribution ${\rm P}(m,k)$ is assumed to be uniformly distributed. \par
In this study, on the basis of the framework of Bayesian model selection and empirical Bayesian methods\cite{bishop2006pattern}, the DNN feature $m$, the kernel function type $k$, and its hyperparameters $\theta_{\rm k}$ were determined by maximizing the following marginal likelihood functions called Bayesian evidence: 
\begin{eqnarray}
&\:&{\rm P}(m, k, \theta_{\rm k}|{\mathbf Y}) \propto {\rm P}({\mathbf Y}|m, k, \theta_{\rm k}) 
= \int d {\mathbf F} {\rm P}({\mathbf Y}|{\mathbf F},m, k, \theta_{\rm k}) {\rm P}({\mathbf F}|m, k, \theta_{\rm k})  \\
&=&\exp\left\{\frac{\beta}{2}\mathbf{\mu'}^T\Lambda_{11}^{-1}\mathbf{\mu'} - \frac{\beta}{2}\mathbf{Y}^T \mathbf{Y}\right\}\left(\frac{\beta}{2\pi}\right)^{N/2}\left(\frac{1}{2\pi}\right)^{d/2}\\
&\:&\:\:\:\:\:\:\:\:\times\int d\mathbf{F} \exp\left\{-\frac{\beta}{2}\left(\mathbf{F} - \Lambda_{11}^{-1}\mathbf{K}\mathbf{Y}\right)^T\left(\mathbf{K}^{-1} + \beta I\right)\left(\mathbf{F} - \Lambda_{11}^{-1}\mathbf{K}\mathbf{Y}\right)\right\} \\
&=&\exp\left\{\frac{\beta}{2}\mathbf{\mu'}^T\Lambda_{11}^{-1}\mathbf{\mu'} - \frac{\beta}{2}\mathbf{Y}^T \mathbf{Y}\right\}\left(\frac{\beta}{2\pi}\right)^{N/2}|\Lambda_{11}^{-1}|^{1/2}.
\end{eqnarray}
In the Bayesian model selection framework, the likelihood of a model or hyper-parameter is often compared not with the marginal likelihood itself, but with its negative logarithm called Bayesian free energy\cite{watanabe2018mathematical},
\begin{eqnarray}
F(m, k, \theta_{\rm k}) = -\log[p(m, k, \theta_{\rm k}|\mathbf Y)],
\end{eqnarray}
where $k$ denotes the kernel type. 
This Bayesian free energy is also employed in this study. \par
Since $\theta_{\rm k}$ of the kernel function has continuous values, it is not possible to search for the value that minimizes the Bayesian free energy in a grid search like $m$ or $k$. 
Therefore, in this study, the L-BFGS-B algorithm~\cite{liu1989limited}, a kind of quasi-Newtonian method, is used to find $\theta_{\rm k}$ that minimizes the Bayesian free energy. 
Newton's method becomes a local solution depending on the initial values of $\theta_{\rm k}$. 
In this study, we explored the initial values that minimize $F(m, k, \theta_{\rm k})$ in the range shown in Table~\ref{tbl1}. 
The model parameters $k$ and $m$ were determined with $\theta_{\rm k}$ optimized in this manner.

\bibliography{main}
\bibliographystyle{junsrt}

\end{document}